# *On the effect of strain and triaxiality on void evolution in a heterogeneous microstructure – A statistical and single void study of damage in DP800*

Carl F. Kusche[1], Felix Pütz[2], Sebastian Münstermann[2], Talal Al-Samman[1], Sandra Korte-Kerzel[1]

[1]Institute for Physical Metallurgy and Materials Physics, RWTH Aachen University, 52062 Aachen, Germany

[2]Steel Institute, RWTH Aachen University, Intzestrasse 1, 52074 Aachen, Germany

## Abstract

In order to improve the understanding of damage evolution in mechanically heterogeneous microstructures, like the ones of dual-phase steels, the influence of the applied stress state is a key element. In this work, we studied the influence of the globally applied stress state on the evolution of damage in such a microstructure. Classical damage models allow predictions of damage during deformation based on considerations of the material as an isotropic continuum. Here, we investigate their validity in a dual phase microstructure that is locally dominated by its microstructural morphological complexity based on a statistical ensemble of thousands of individual voids formed under different stress states. For this purpose, we combined a calibrated material model incorporating damage formation to assess the local stress state in samples with different notch geometries and high-resolution electron microscopy of large areas using a deep learning-based automated micrograph analysis to detect and classify microstructural voids according to their source of origin. This allowed us to obtain both the continuum stress state during deformation and statistically relevant data of individual void formation. We found that the applied plastic strain is the major influence on the overall number, and therefore the nucleation of new voids, while triaxiality correlates with the median void size, supporting its proposed influence on void growth. In contrast, coalescence of voids leading to failure appears related to local instabilities in the form of shear band formation and is therefore only indirectly determined by the global stress state in that it determines the global distribution, density and size of voids.

## 1. Introduction

Dual-phase steels, as a widely applied variety of advanced high-strength steels (AHSS), have yielded a high research interest for many years, especially focusing on their deformation behaviour [1-4] and formation of microstructural damage [5-7] in the form of deformation-induced voids. Due to the contrast in mechanical behaviour between their constituting phases, ferrite and martensite, a complex partitioning of stresses and strains takes place. This behaviour has been analysed by both simulative and experimental approaches [8-10].

The micromechanical mechanisms of void initiation are commonly identified as a brittle cleavage fracture of martensite islands (martensite cracking), in contrast to plastic decohesion processes of martensite-ferrite interfaces and ferrite-ferrite grain boundaries [6, 11]. However, which of these mechanisms contributes more to damage and eventually material failure, will depend mainly on the type of considered dual-phase microstructure, stress conditions and applied strain [8, 11, 12].

Modern in-situ deformation devices operated inside high-resolution scanning electron microscopes (SEM) are able to visualize these micromechanical mechanisms of void initiation as well as the development of voids during deformation [13], but under an altered stress state due to the free surface in an in-situ experiment. Additionally, the intrinsic microstructural heterogeneity of commercial dual-phase steels in terms of phase morphology, distribution and density at both the µm and mm (sheet) scale leads to a dominance of the local, microscopic stress state, caused by the individual morphology of the local microstructure on the initiation and growth of voids at each site.

Consequently, experiments proving the dependence of damage in the bulk material on stress triaxiality in the intrinsically heterogeneous dual-phase microstructures are more challenging. Although the effect of triaxiality on damage evolution is believed to result predominantly in the growth of voids, experimental studies largely relied on model materials with simple microstructures to ensure a homogeneous stress distribution along the samples [14, 15]. However, due to the individual conditions of the local microstructure, and therefore stress, strain and deformation conditions [4, 16, 17], as well as the typically very heterogeneous distribution of phases in commercially applied dual phase steel sheets, statistical information gained through high resolution inspection of large deformed areas is vital to accurately assess the magnitude of introduced damage [11, 18].

Recently, automation based on neural networks has been applied to enhance segmentation of microstructural features in these types of steels [19] as well as to classify different types of damage events in order to obtain significantly larger amounts of statistical data on operative damage mechanisms [12]. Applying such automated recognition and analysis tools for deformation-induced voids is advantageous in two ways:

(i) It enables a simple, statistical approach on large areas to reveal global quantities, such as a damage variable in the sense of e.g. a void area fraction.

(ii) It regards every emerging microstructural void individually, so that individual measurements of damage parameters like void size, morphology and mechanism of initiation are obtainable.

In this way, we employ statistics of individual voids and correlate these to stress and strain parameters in order to gain information about the fundamental mechanisms of void formation stages, classically identified as void initiation, growth and coalescence [20].

In theoretical approaches [21], stress-state dependent factors have been proposed, in particular a direct influence of stress triaxiality on the intensity of damage. Experimentally, this dependence could be verified using both SEM-based [22] and 3D-microtomography methods [15]. However, these connections are much more challenging to identify when dealing with real microstructures that typically show a significantly altered and locally heterogeneous stress and strain due to their microstructure [13]. This microstructural heterogeneity has been proven to lead to a significant inhomogeneity in spatial damage distribution [18] and development of apparently similar individual damage site in an in-situ experiment observing the material surface [12]. Therefore, expected correlations between damage and the continuum stress state would only be expected to match experimental data in a statistical way.

In this work, we employ large-area void statistics to reveal the convergence of the damage statistics in a heterogenous and complex dual-phase microstructure against the theoretical expectations for the stress state influence expected for isotropic, homogeneous materials in a statistical ensemble of thousands of voids. In this way, the widely-applied models of damage evolution [23] are appraised with respect to the postulated causes for void initiation and growth that can now be tracked and attributed to specific parameters of stress and strain through the application of statistical evaluation in a real microstructure. For the third fundamental mechanism of damage formation in dual-phase steels, void coalescence, preceding stages of void nucleation and growth have been observed prior to failure of the sample [24], and attributed to the existence of shear bands [7].

In order to understand the failure behaviour of the deformed material, it is crucial to understand the complex interplay between the aforementioned mechanisms. De Geus et al. [25] reported that loading conditions, as well as microstructural factors like mechanical contrast between the constituting phases, volume fraction and morphology exert a strong influence on the failure behaviour. Considerable theoretical and experimental efforts to model the failure process were made by Ramanzani et al. [20], who however did not take the formation of deformation-induced voids into account. While an interaction between the introduced voids and the process of sample failure has been described earlier [7], the global patterns and dependencies of these interactions have yet to be unravelled. This is particularly important due to the necessary large-scale, high-resolution observations. Equally, extensive advances have been made in modelling deformation and damage initiation processes on the microscale [9, 24]. Typically, simulative approaches to these processes rely on local microstructural data around the considered individual sites of damage initiation or evolution, leading to insights about microstructurally favoured sites for void initiation [26] or plastic deformation [4].

To deliver reliable data about the local stress state during deformation, a calibrated FEM material model is usually applied, in this case with the additional incorporation of a damage criterion. In the field of damage mechanics, two different types of models exist: Coupled and uncoupled models [27]. For coupled models, a damage variable is applied to reduce the yield potential in accordance with the material's softening, induced by ductile damage. A typical example of this kind of model is the Gurson-Tvergaard-Needleman (GTN) model [28-30], which is an example of micromechanical models that can account for physical behaviour of damage evolution by their sets of parameters. Since these parameters are independent, an extensive iteration process is required for their calibration [31]. As a complementary approach, uncoupled models are widely used and describe the material's behaviour without considering softening by ductile damage, like the Bai-Wierzbicki (BW) model [32].

The current study comprises three major parts: In the first part, we examine the damage intensity in the form of void area fraction as a function of different stress - strain conditions and establish respective correlations between these parameters. In the second part, the collected individual damage data is used to trace back the initiation and growth of voids to the triaxiality parameter through the advantage of statistical investigations. In the third part, the final stage of damage formation, i.e. void coalescence, is investigated by analysing characteristic microstructural patterns such as shear bands formed during deformation and clarifying its interaction with void formation before fracture. By considering these three aspects, we aim to deepen the understanding of global damage formation processes and bridge the gap between isotropic behaviour of single voids in model materials one the one hand and the integral behaviour of an ensemble of voids in complex microstructures on the other.

## 2. Experiments and Methods

### 2.1 Tensile testing

In this study, a commercial dual-phase steel of DP800 grade, manufactured by thyssenkrupp AG, was used in the form of sheet metal with a thickness of 1.5 mm. Various geometries of tensile samples, leading to altered stress states in the gauge part, were developed and manufactured using electric discharge machining (EDM). These geometries were used to calibrate the material model to the obtained force-displacement data in order to have good agreement between experiment and FEM-simulations. The geometries of the notched samples are depicted in Figure 1. They can be divided into two groups: (a) Notched geometry over the sheet thickness (Figure 1(a)), used to realise stress states close to plane strain condition and (b) other notched geometries with lateral notches (Figure 1 (b)). All samples were deformed with the tensile axis parallel to the rolling direction. Additionally, a simple tensile geometry (without a notch) was tested to fit the flow curve of the material (Figure 1(c)).

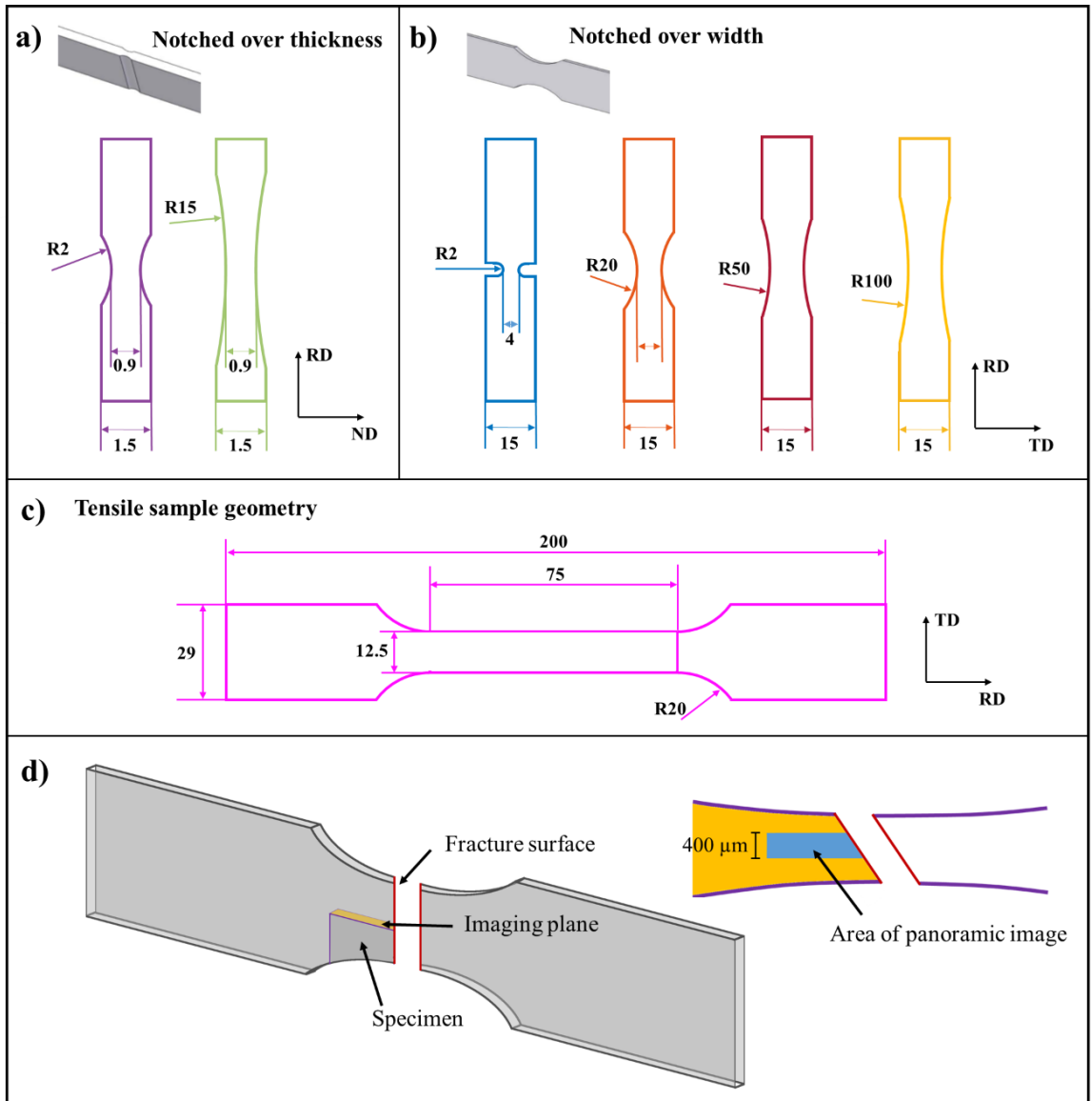

Figure 1: Tensile geometries used in this study. a) plane-strain geometries notched over the sheet thickness (width 20 mm), b) notched tensile geometries notched over sample width (width 12.5 mm), c) simple tensile geometry without notch. The color-coding for sample geometries is consistent with all following figures. All measurements are given in mm. d) Illustration of specimens cut from tensile samples and area of the evaluated panoramic image.

After deformation to fracture at a rate of 0.005 min$^{-1}$, specimens were cut from the tensile samples as shown in Figure 1 d. The plane marked in yellow was evaluated via SEM panoramic imaging within an area of 400 µm width (Figure 1d). Samples were ground up to 4000 grit, mechanically polished (6, 3, 1 µm & colloidal silica suspension (OPS)) and subsequently etched in 1% Nital solution for 5 s. The images were acquired from samples deformed to fracture on a plane perpendicular to the transverse direction. As commercial DP steels often exhibit a banded microstructure, an imaging plane perpendicular to the transverse direction was chosen. In this way, the obtained panoramic images contain multiple bands in favour of several large regions of varying martensite content.

## 2.2 FE-Simulations

This study utilizes the modified BW (MBW) model, specifically the np-MBW-19 version [33]. This version features damage initiation, as well as ductile fracture loci, that specify the material's damage and fracture behaviour for different stress states. Additionally, an indicator is applied to account for non-proportional loading during the deformation of the material. For further information about the applied material model, we refer to previously published studies[31, 33, 34]. To use the material model in this context, it is important to perform a thorough calibration. Therefore, tensile tests of different geometries were conducted to account for different stress states. Afterwards, simulations were carried out using a VUMAT subroutine, which implemented the np-MBW-19 model into ABAQUS. The conventional tensile test was used to fit the flow curve of the material, whereas the other tensile tests, described in Figure 1, were employed to account for the damage initiation and ductile fracture loci. The determined material parameters for the steel used in these investigations were published in an earlier study on damage in dual phase steels [33]. After the calibration process, the FE simulations were able to provide an accurate representation of the local stress state in the sample. Thus, all experimentally tested samples were simulated and their stress triaxiality, as well as plastic equivalent strain were evaluated in the last step before simulated sample fracture. Subsequently, stress triaxiality and plastic strain were extracted at the nodes of the mesh according to the areas, where images were acquired, as described in the previous chapter.

## 2.3. Void Recognition and Analysis

The automated void detection and classification approach discussed in [12] was applied on panoramic SEM images acquired using secondary electron detection (Zeiss LEO 1530 FEG-SEM and FEI Helios NanoLab 600i). All images were taken as individual frames of 100 µm horizontal width, corresponding to 3072 pixel (px) with 20% overlap to the next frame. The single images were stitched into panoramic images using a MATLAB-based site recognition and stitching algorithm based on VLFeat toolbox [35].

Voids emerging in micrographs of deformed dual-phase steel samples can have various origins. In particular, they may be distinguished as deformation-induced or not, such as inclusions. In a polished cross-section, the latter can either be due to particles leaving the surface during metallographic preparation or remaining inclusions appearing black in the SEM micrographs. To detect these sites and exclude them from the obtained statistics of deformation-induced damage, a neural network further described in [12] was applied. It is noted that the method is able to detect and filter out both types of inclusion-related voids, i.e. those that are still present in the microstructure and others that have fallen out during sample preparation. The accuracy of this procedure is usually larger than 95% [12]. In this method, detection of voids was carried out by a grayscale thresholding and clustering via a DBScan algorithm [36]. Each individual void was then passed over to a convolutional neural network (Inception V3 [37]), that had been trained on manually labelled voids to distinguish deformation-induced voids from voids introduced to the micrographs by inclusions or other artefacts. Deformation-induced voids were processed further via a watershed algorithm, in order to obtain statistical information on individual void area and void shape that can be used for further analysis. The approach is shown schematically in Figure 2 for an exemplary void in an SEM-panorama.

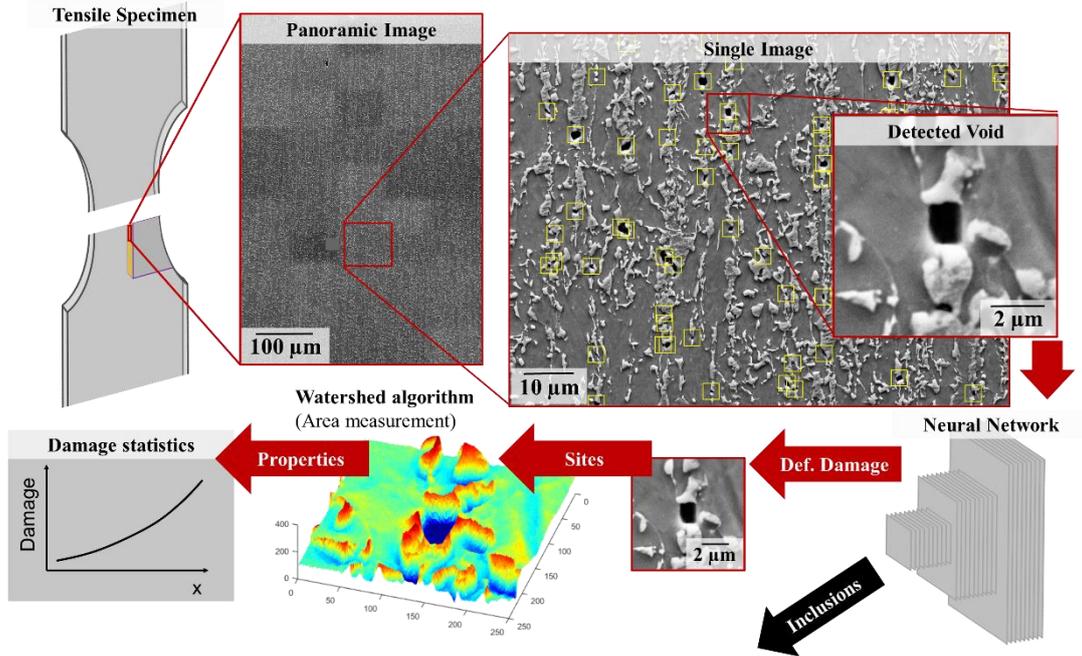

Figure 2: Experimental concept of automated void detection and classification in panoramic SEM images using a neural network to distinguish deformation-induced voids from inclusions.

Void area fractions were obtained using vertical binning along the tensile axis, with every data point representing the area fraction of damage voids in a sample area as wide as the complete panoramic image (400 µm) and 3000 px (97.66 µm) in height. The spacing between the bins was 600 px (19.53 µm).

## 3. Results

### 3.1 Morphology of voids

The detected voids on the panoramic micrographs showed voids that can, in analogy to classifications found in the literature, be classified into three main categories:

(a) Freshly nucleated voids that clearly reveal the underlying mechanism of initiation.

(b) Voids that have grown with further plastic deformation.

(c) Significantly larger voids close to the fracture surface, which seem to have originated from the coalescence of several pre-existing, deformation-induced voids.

In addition, inclusions were equally detected as voids, but separated from the deformation-induced voids by the use of the aforementioned neural network.

As seen in Figure 3, the two main damage mechanisms found in the DP steel used in this study were martensite cracking and the decohesion of martensite/ferrite interfaces. Figure 3 shows the fundamental stages of damage formation detected in the deformed microstructure, namely initiation (a+b), void growth (c+d) and coalescence (e). All voids are expected to undergo these three stages of void evolution irrespective of initiation mechanism, i.e. martensite cracking (a+c) or martensite/ferrite interface decohesion (b+d). In addition to the deformation induced voids, typical inclusions are depicted in Figure 3 (f+g), which were detected by the neural network and not processed further.

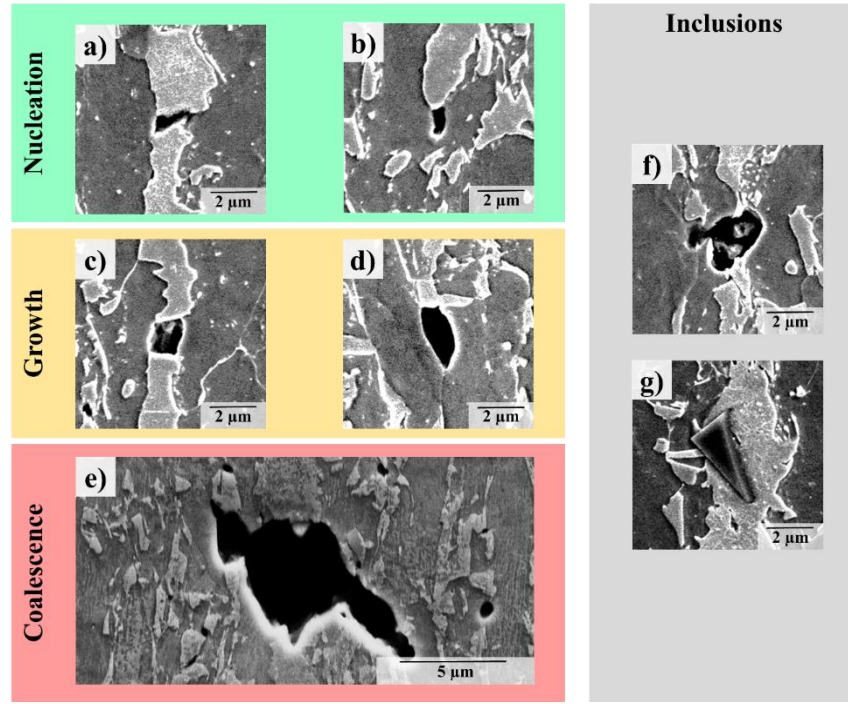

Figure 3: Classification of observed microstructural voids in deformed samples. a) martensite cracking, b) decohesion of M/F interface, c) grown martensite crack, d) grown interface decohesion site, e) coalescence of voids close to fracture surface, f) site of inclusion fallen out of the sample surface during preparation classified by neural network, g) inclusion (likely TiN) in a martensite island classified by neural network.

## 3.1 Influence of stress state on void area fraction

The samples utilized for this study were designed to cover a wide range of stress states from uniaxial tensile stress to plane strain conditions. For the FE simulations of the uniaxial loading the samples were exposed to, the measured displacement was used to simulate the local stress and strain conditions during deformation up to the point of sample fracture in the experiment. Stress triaxiality and strain were found to monotonically increase over the process of deformation, so that the highest value for these quantities and the value at the point of sample failure coincided. Therefore, the last timestep at sample fracture was used for evaluating the stress and strain state in order to correlate with the microstructural damage measurements. All considered points per sample are spatially distributed along the tensile axis in the middle of the sampled area, as shown in Figure 4.

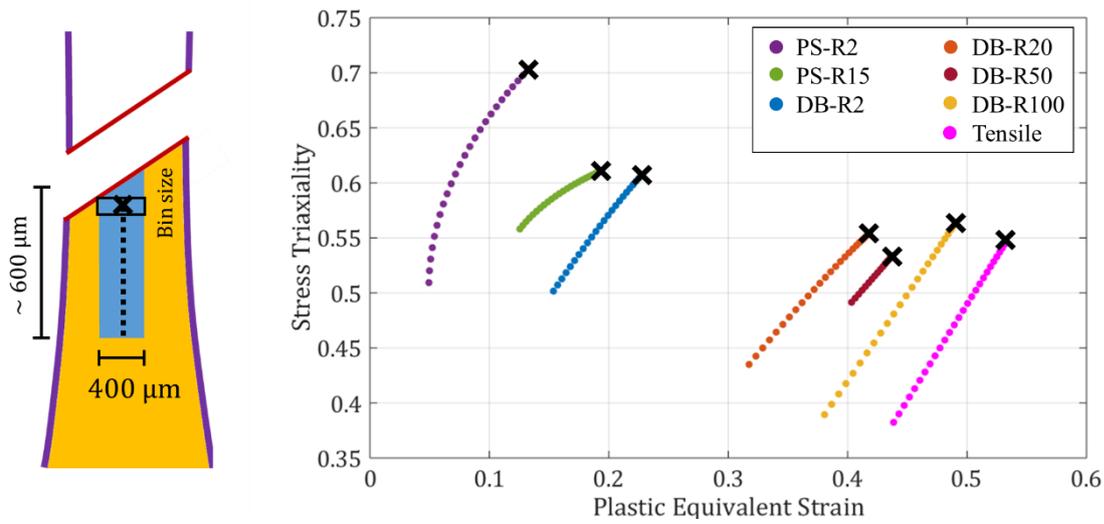

Figure 4: FE simulations of triaxiality-strain date for the investigated tensile samples at a timestep coinciding with sample fracture. Each data point represents one measured and evaluated rectangular bin from the area segment in the SEM (blue sample area, bin indicated by thin black line). The cross points in the diagram corresponds to the cross point position marked on the sample directly below the fracture surface.

As visible in Figure 4, the covered window in stress triaxiality and plastic equivalent strain ranges from 0.37 to 0.71 in triaxiality and 0.05 to 0.55 in plastic strain. Only positive values for triaxiality were considered, as the most common models of void growth only apply to these values. For higher stress triaxiality values, the plastic strain at fracture was lower.

Area fractions of detected deformation-induced voids along the tensile axis of the samples were calculated for up to 600 µm from the centre of the fracture surface with respect to the tensile axis as illustrated in Figure 4. As the spatial evaluation points in the FE simulation and the panoramic image were not the same, a polynomial interpolation was used on the FE data to approximate values for the exact locations of the evaluated bins in the panoramic image and obtain an identical spatial grid for evaluation. The values obtained this way for the global damage variable of void area fraction were compared with the plastic equivalent strain for the same point in the sample. The results are shown in Figure 5.

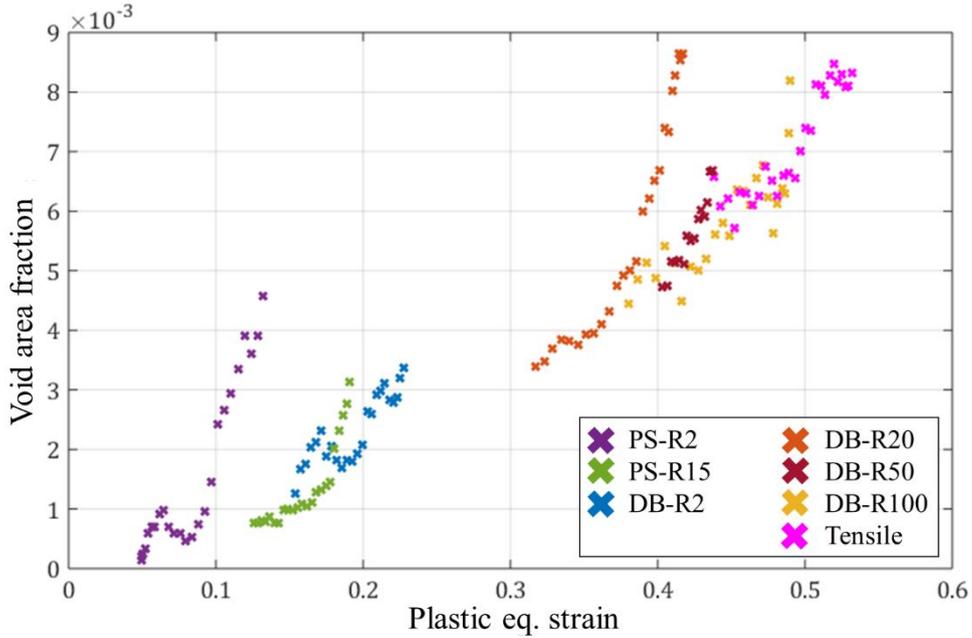

Figure 5: Measured area fraction of voids vs. estimated plastic equivalent strain from FE simulation.

The data reveals a global trend of increasing void area fraction with increasing strain. However, further examination of the data reveals additional insights: (i) Damage formation is found to increase at different rates for the different stress and strain states investigated. (ii) The overall range of strains varies strongly with the different sample geometries investigated. (iii) Even though all samples were measured at the point of maximum deformation at fracture, the values of final void area fraction are different.

The imposed stress and strain state through the altered sample geometry was investigated further to elucidate the influence of stress triaxiality on the measured void area fraction (Figure 6).

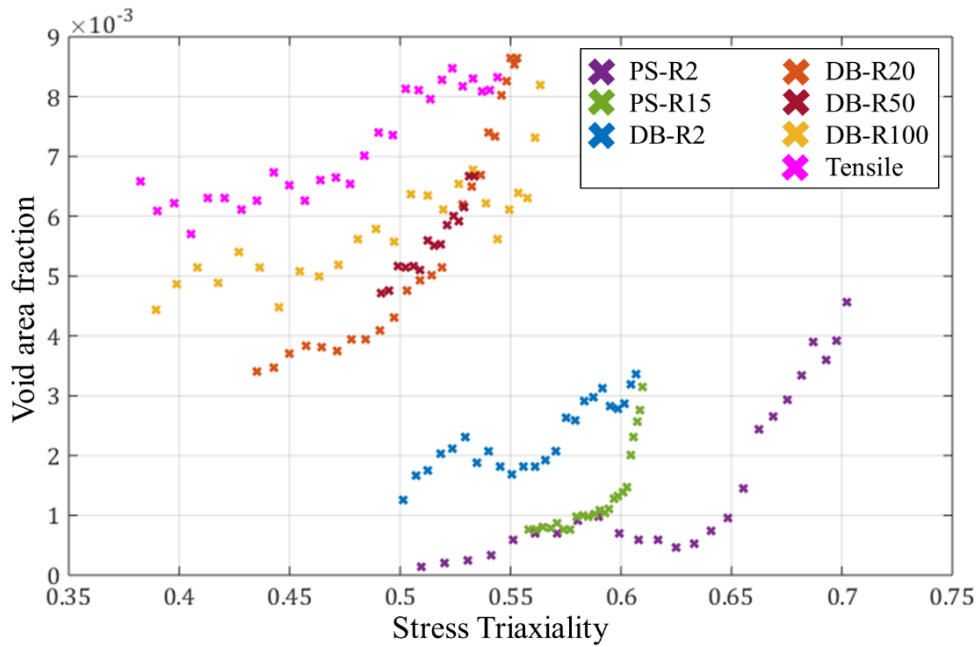

Figure 6: Measured area fraction of voids vs. estimated stress triaxiality from FE simulations.

The imposed stress and strain state through the altered sample geometry was investigated further to elucidate the influence of stress triaxiality on the measured void area fraction (Figure 6). A comparison across all the cases reveals that large void area fractions populate the upper-left corner of the plot associated with lower triaxiality and lower void area fractions are associated with the higher values of stress triaxiality. However, when considering each tensile geometry individually, an increasing void area fraction with increasing triaxiality is apparent.

### 3.2 Influence of plastic strain and stress triaxiality on void size distribution

While the statistics gathered from the large-scale observations yield a clear and distinct picture regarding the influence of stress triaxiality and plastic strain, all presented data to this point refers to the cumulative property of void area fraction. To gain more insights into the stages and mechanisms of void formation, information about every single of the ten thousands of detected and measured damage sites was evaluated. This enables us to also consider void size distributions in addition to area fractions or number of sites as shown in Figure 7.

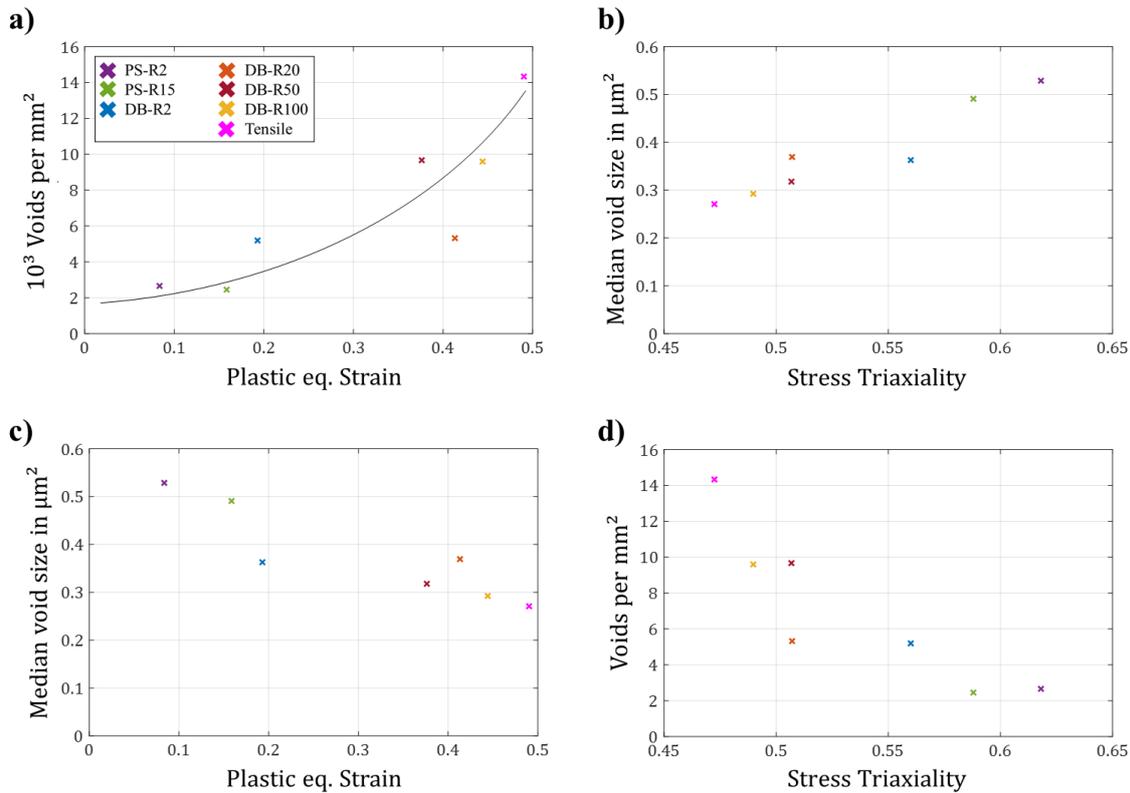

Figure 7: Number and size of voids as a function of strain and triaxiality. a) Plastic equivalent strain and number of deformation-induced damage sites per area, b) integral stress triaxiality over the evaluated area and median void size c) plastic equivalent strain and mean void size, d) integral stress triaxiality over the evaluated area and number of detected damage-induced voids per area. For the data in plotted in (a) an exponential curve has been observed elsewhere and is therefore shown here as a fit.

Again, trends are visible in this data: While the number of voids emerging in the microstructure increases with the calculated plastic strain, the opposite trend is seen in Figure 7c for the median void size. For the first an exponential fit is also shown as this is the distribution found in previous work on the same material for the number of voids per area as a function of uniaxial strain [12]. A relationship more adequately described as a linear increase can be found between stress triaxiality and the median void size (Figure 7b), whereas in Figure 7d the number of damage sites dropped with increasing triaxiality. For these data points shown in Figure 7, all values from the FE simulations were obtained as an average of the evaluated area, thus the area of the panoramic image was evaluated as a single bin to obtain one integral value of stress triaxiality per sample geometry.

Even though different strains lead to rupture of the sample, as seen in Figure 4, and growth of voids is more pronounced in sample geometries with lower fracture elongations and therefore high triaxialities, it is noted that the levels of damage at the point of sample failure are by no means equal, as seen in Figure 6. This material failure, however, is caused by the third stage in the process of void evolution, the coalescence of voids. This has been investigated using the large-scale images of the deformed microstructure close to the fracture surface.

3.3 Influence of localised shear on the coalescence of damage

Void coalescence is the final stage of damage formation before failure occurs. From the obtained panoramic SEM images, statistical information regarding the location, size and morphology of voids have been collected in order to study the spatial patterns of void coalescence. This was done to elucidate whether they depend on the interaction of local deformation with existing voids. By focusing on the shape of the voids, especially their angle towards the tensile axis, it was found that large voids, having originated from multiple initiated voids by coalescence, are often oriented just under 45° to the tensile axis as shown in Figure 8.

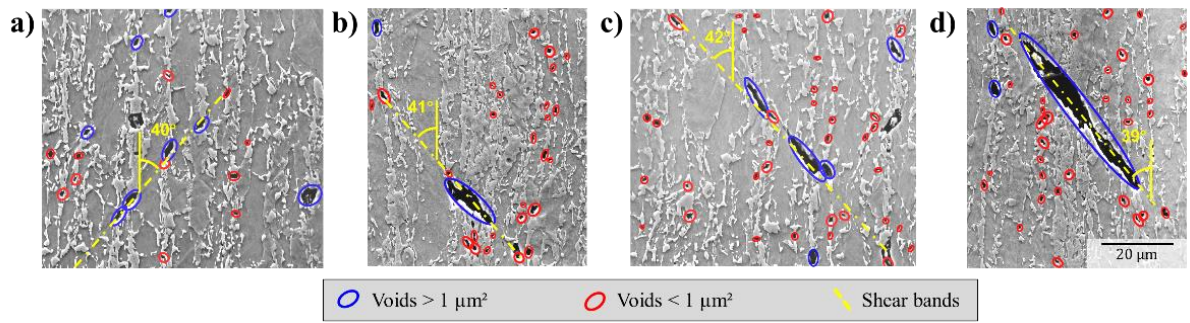

Figure 8: Extracted regions of interest from panoramic images showing various stages of void coalescence. a) the initial formation of elongated voids along a shear band, b) the coalescence of voids, forming one large void in a largely martensitic region and c) regions of average phase distribution. d) the formation of an even larger void near the sample surface through further void coalescence.

Whether these voids have entered the stage of void coalescence was determined by regarding the adjacent martensite islands: a grown void is thought to typically still lie in between or at the interface of a single martensite island, while void coalescence is understood as the joining of voids originating from more than one interface / martensite island. For these voids, orientation angles just below 45° to the tensile axis could predominantly be observed. Furthermore, the large area observations have yielded additional information about the patterns of voids: The detected, angled voids tend to form along straight lines with the inclination of the voids along "coalescence bands". Figure 8 b shows an example of primary coalescence, initially joining voids together. This is typically found in the widest martensite bands. Adjacent to these sites, along the slope of the inclination angle of the resulting coalescence site, the microstructure appeared to be sheared by slip bands, which promotes additional coalescence of pre-existing voids, leading to multiple coalescence sites along the shear bands as shown in Figure 8 c) and finally, secondary coalescence of those, as depicted in Figure 8 d.

## 4. Discussion

The three stages of void formation in a highly heterogeneous dual phase steel microstructure were investigated in a statistical way using automated imaging, void recognition and deep-learning based analysis. In this way, this study enabled us to gather a large amount of data from > 10,000 voids in order to investigate the void formation behaviour in these complex microstructures and correlate the findings to simulations of the stress and strain state in the deformed samples. We will discuss these here starting with the global picture of median void size and area fraction before considering their distribution in more detail.

## 4.1 Quantitative dependence of global damage accumulation on strain and triaxiality

The data in Figure 5 shows a consistent influence of the applied strain on damage evolution with the void area fraction increasing with strain. In contrast, the impact of stress state (triaxiality) is less obvious from the data in Figure 6. Here, we observe (i) for each sample an increase in void area fraction with stress triaxiality, however, globally we find (ii) lower void area fractions with decreasing triaxiality. The first case, considering each sample individually, is in agreement with the analyses by Lemaitre, Gurson and McClintock [21, 38, 39]. The observed inconsistency between the findings for each single specimen and the global trend then needs to be considered in more detail.

As the correlations in Figure 5 and Figure 6 implicitly contain the relationship between strain and triaxiality values for the examined specimens (Figure 4), the dependences of void area fraction on the plastic strain and stress triaxiality are better considered simultaneously (Figure 9) to avoid overlooking correlations intrinsic to the chosen samples between these values. For each sample geometry and therefore curve, different levels of plastic equivalent strain are present, but an increase in void area fraction is nevertheless visible for both parameters, strain and triaxiality. For an easier observation of the trends depicted in Figure 9, the graph shows a surface fit based on the data points.

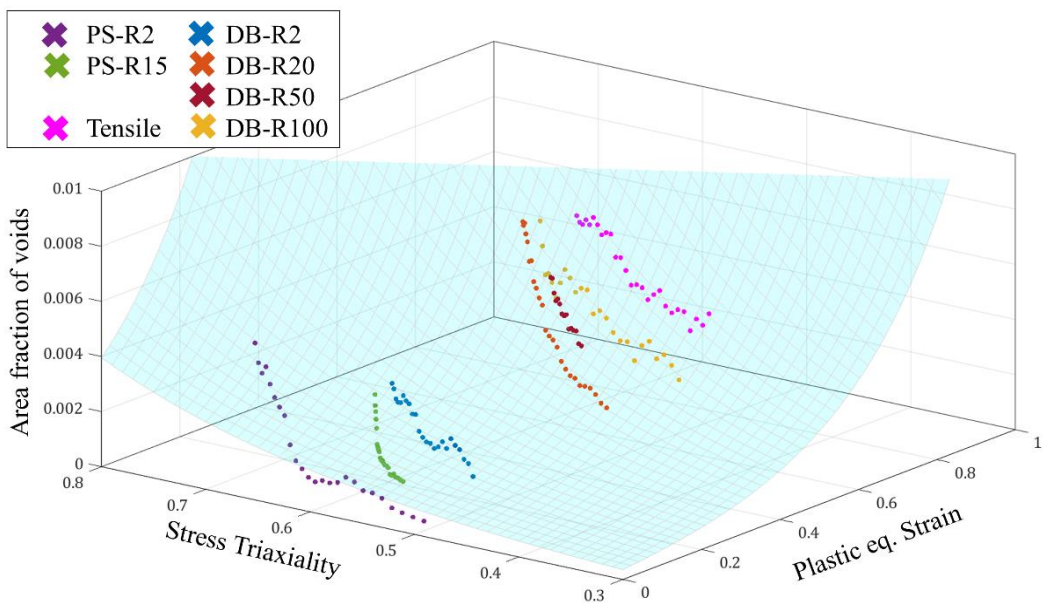

Figure 9: Evolution of void area fraction as a function of stress triaxiality and plastic equivalent strains from FE-simulations. The blue area represents a double exponential surface fit of the data points.

We do note, however, that triaxiality is calculated as the ratio of the hydrostatic stress and the von Mises equivalent stress and is therefore in itself only a measure of how dominant the hydrostatic stresses are in the applied stress state. The triaxiality does not, however, take the overall magnitude of the applied stress into account. Here, we evaluated all samples at the point of failure and within 600 µm of the fracture surface, where they had, presumably, all reached a consistent true stress level. If void area fractions were additionally to be considered for samples before fracture or far away from the fracture surfaces, the results might therefore differ if only strain and triaxiality were to be compared. In such a case, the use of the hydrostatic stress magnitude may be more accurate when comparing samples of varying states of deformation.

In principle, our observations therefore confirm the expected evolution of void density and area with respect to plastic equivalent strain and stress triaxiality in a material that shows a high internal mechanical contrast between the two constituent phases and also inhomogeneity (martensite bands) across its thickness. Previously, this trend had also been confirmed by others [7, 14, 40] using large area or volume measurements, but normally at a lower spatial resolution.

The classical expectation [21] is to not only find an increase in cumulative damage quantity with triaxiality, but in particular growth of voids. To achieve this correlation, the influences of strain and stress triaxiality have to be separated and discussed concerning other measured quantities of the void ensembles, rather than discussing area fraction as a cumulative property.

### 4.2. Separation of the stages of void evolution for individual voids

It is perhaps not surprising that, once voids have reached a size that encompasses several instances of each of the phases of the dual phase steel, the behaviour of homogeneous model materials is reproduced. Here, we therefore set out to investigate whether this behaviour is in fact observed at the smallest observable scale as well, where voids contained entirely inside a martensite island or ferrite grain can be resolved. This implies that we need to follow previous attempts to distinguish between those damage sites which have been freshly nucleated at a single interface, grain boundary or as a crack across a martensite island and those which have already grown significantly.

For this, a threshold-based criterion using a fixed threshold has been proposed for SEM-based methods and for the purpose of parameter fitting in a GTN-type model by [7, 23]. Establishing a lower bound threshold for grown voids then requires a statistical approach to damage, which is possible based on SEM data or X-ray microtomography [41]. A main difference between methods then lies in the expected resolution and therefore also the length scale of possible thresholds. For 3D-synchrotron observations, Landron et al. [42] determine that with a voxel size of as low as 100 nm, a significant effect of voids with a diameter smaller than 4 µm is observed. In contrast, the 2-dimensional pixel size using SEM is much smaller, e.g. 32 nm in this study, and voids as small as 0.02 µm² can therefore be detected and considered. Avramovic-Cingara et al. [40] used a threshold of 1 µm in diameter to distinguish between nucleated and grown damage sites.

However, a comparison of our measurements of detected voids with respect to their size in different stages of their development show that these threshold-based criteria prove challenging to set up in a physically meaningful way.

The size distribution of voids imaged and analysed across all samples in this study is shown in Figure 10 along with several micrographs showing all three stages of damage colour-coded as (close to) nucleation (green), individual void growth (yellow) and coalescence (red). A direct comparison may be drawn with the threshold of 1 µm suggested by Avramovic-Cingara et al. [40] for micrographs of similar resolution on DP600. In their work, the authors found an average aspect ratio of 2, giving an equivalent area threshold of the order of 0.5 µm². A comparison with the void size distribution in Figure 10 shows that in our material, the majority of damage sites is in fact smaller and therefore below this threshold.

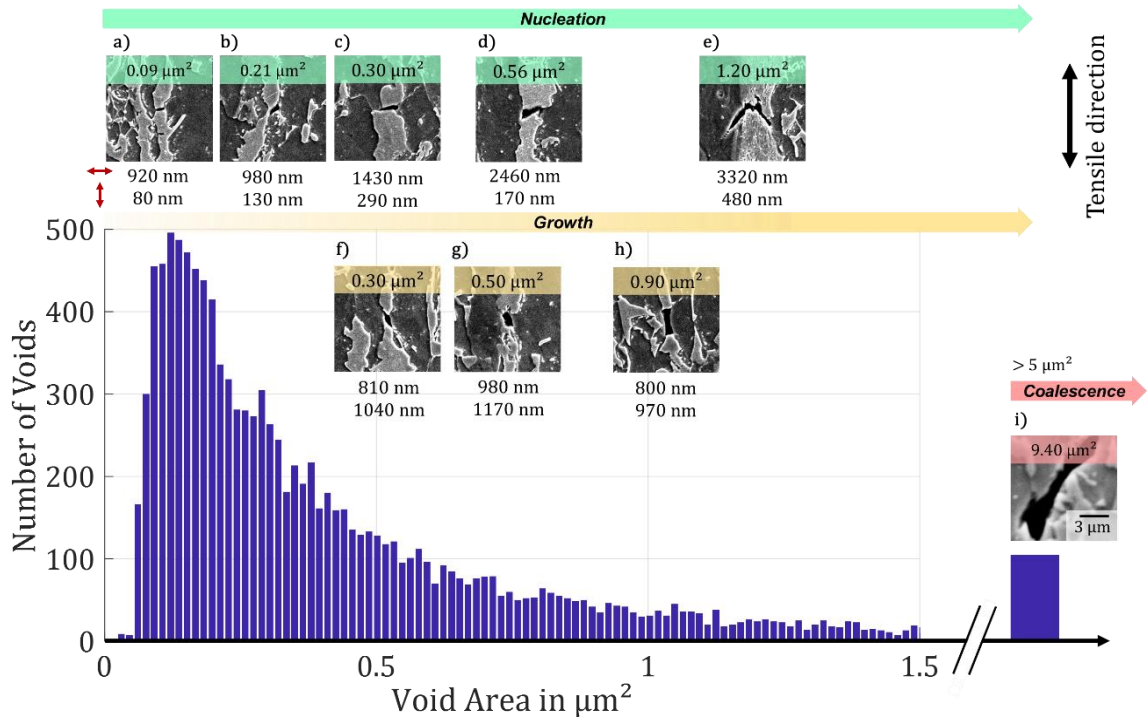

Figure 10: Void size distribution of all detected voids (12730 sites total), voids larger than 5 µm² marked with a single bar and corresponding example of a void coalescence site. a)-i) Examples of voids showing various stages of nucleation, growth and coalescence. Measured values are given as the void width and length of growth in tensile direction.

Figure 10 (a) and (b) include voids close to nucleation that are associated with the smallest measured areas for martensite cracking (a) and interface decohesion (b). As interface decohesion may be more easily pictured as a rather continuous process once a void has nucleated, the remaining voids selected for Figure 10 are martensite cracks, starting and ending at the surfaces of a martensite island. Owing to this geometry, it is possible to determine both the initial crack length, given by the length of the visible crack surface for a through-thickness crack of the martensite island, and the displacement during void growth, given by the distance between the crack surfaces. The displayed voids highlight the independence of the state of void evolution on a void's area: A similar void area can be formed in a crack with a large crack length but small crack opening or a short crack showing extensive growth. For example, in comparison of Figure 10 (e) and (h) a crack opening larger by a factor of nearly 7 in (h) compared with (e) still results in a smaller void area. This can easily be understood by comparing the width of the cracked martensite islands that measure 3320 nm (e) compared with 800 nm (h). Similar examples are given in Figure 10 with two voids of 0.30 µm² (c and f) and around 0.50 µm² (d and g). Each pair possesses very similar areas but, when analysed individually with respect to their width, displacement and surrounding microstructure, show different stages of evolution. In the case of (a) the small area is due to both a small crack opening and the small lateral extension of the martensite island. The entered stage of void evolution therefore cannot be determined by comparatively simple measurements of void area or aspect ratio alone. For martensite cracks, further factors may help in a better discrimination, including crack opening distance, orientation to the main stress axis or relation to the size and shape of the original martensite island. In the case of interface decohesion, a more continuous process may be envisaged once a void has formed and a physical threshold for the first decohesion process may be underneath that resolved routinely by SEM.

In any case, a statistical evaluation of the different stages of damage formation with a clear separation of the stages will require correlation of damage mechanism, surrounding microstructure and geometric parameters for every given void and a clear working definition of where nucleation has completed and void growth begins. Once significant growth has taken place, detection and determination of geometric parameters of the void become easier, but conversely, the identification of a single underlying or damage mechanism or origin of the void in terms of a nucleation mechanism become much more difficult to achieve. This is true both for a manual analysis as mechanisms begin to interact and origins become obscured and as a result also for automated analyses based on artificial intelligence that rely on the first manual analysis for their training [12].

In summary of the above, nucleation of new voids in the sense of an increase in void number and growth in terms of an overall increase in void area correspond well with those expected based on classical models, even for the strongly heterogeneous dual phase microstructure. However, a clear identification and physical interpretation of nucleation and growth stages for a large number of individual voids will require further progress in high resolution damage characterisation and automated analysis of large image datasets.

### 4.3 Correlation of void size statistics with strain and triaxiality

While a separation of the stages of damage evolution may not yet be possible for individual voids, a more in-depth view into the quantitative values of global void area fraction, discussed above, is nevertheless available based on the single void data obtained here.

Atrain partitioning is accepted to be the main cause of void nucleation in dual phase steels [26, 43, 44], and hence the overall number of detected voids is expected to increase with strain, as shown for samples tested in tension by various researchers [12, 23, 45]. Correspondingly, the automated void detection applied in this study found an increase in the void number density with strain that is consistent with an exponential increase (Figure 7a). In terms of growth, the expectation of a pronounced increase in void size also manifests in the data shown in Figure 7b: the median void size increases linearly to about twice the original value across a triaxiality range of 0.47 to 0.62. At the same time, this increase is associated with decreasing plastic strains at the point of area measurements due to the inverse relationship between triaxiality and plastic equivalent strain at failure (Figure 4). This then results in the decreasing median void size with plastic equivalent strain displayed in Figure 7 c. The correlation between triaxiality and void number remains dominated by the decreasing strain with increasing triaxiality, and therefore suggests no measurable influence of triaxiality on void nucleation. As plastic deformation is dominated by shear and normal, rather than hydrostatic stress components, this corresponds to the expectations in a void-free material. However, micromechanical void initiation and growth processes in a complex, heterogeneous microstructure are governed by local stress and strain partitioning, and therefore microstructural processes, such as dislocation motion and accumulation [7, 17]. For this reason, attempts to trace back the growth of voids to the overall applied stress state have predominantly focussed on homogeneous model materials [15]. Here, the obtained statistics from a multitude of automatically detected damage events enable us to come to similar conclusions in a real and strongly heterogeneous microstructure. While predicting the initiation and evolution of voids at a specific location requires detailed information about the microstructural stress state and constituents, our results show that, applied to a large area, the statistics about void numbers and sizes converge against the expectations based on the commonly used theoretical models for the local continuum stress state obtained by FEM.

The presented approach for quantifying microscale damage by means of automated void detection and classification [12] will therefore enable further development of damage models by making more detailed void size and damage mechanism data available for their calibration. This is particularly true for strongly heterogeneous microstructures, for which it is often simply assumed that the classical models formulated based on homogeneous model materials can be directly applied. For the case of DP800 steel, we have shown here that this is in fact the case, but in the same way, the presented approach could now also be easily transferred to many other materials to confirm assumptions about the effect, or its absence, of local strain partitioning on global damage evolution.

## 4.4 Void Coalescence

In addition to the two stages discussed above, void nucleation and growth of individual voids, the last stage of damage evolution remains to considered, in which a transition from single void growth to their coalescence takes place. It is this stage that is ultimately likely to govern failure of a sample or component by more localised mechanisms, for example by shear banding in ferrite which causes distortion and joining of voids that were originally oriented with an inclination towards the tensile direction. Using the large-scale micrographs acquired here, the occurrence of void coalescence events could also be studied. The pattern of large voids in the We find that the process of void coalescence is connected directly to the locally altered stress state in shear instabilities. It is the panoramic views and large number of coalesced voids (100 voids larger than 5 µm² over all samples) that allowed us here to trace the coalescence sites back to the existence of shear instabilities that are best visible at larger scales and show that multiple sites of void coalescence alignalong these bands of locally altered stress state.

Thus, having studied all three stages of void formation in the dual-phase steel microstructure globally on the obtained large-scale micrographseach stage of damage evolution can beconnected with dominant parameters, many of which consistent with expectations from homogeneous model materials (Figure 11). For a single void, the mechanisms that lead to void initiation and growth are strongly dominated by the local microstructure and morphology. However, when regarding all microstructural sites in a statistical way, including large areas and therefore whole ensembles of thousands of voids, the microstructure-independent continuum stress state alone proves sufficient to obtain numbers converging against the expectations for void initiation and foremost, growth in an isotropic material as proposed and shown in [15]. While every single void is locally dominated by its surroundings, the global connection between triaxiality and void growth remains intact. The subsequent the stage of void coalescence, can again only occur where the local boundary conditions in the vicinity of a single void are fulfilled, i.e. a shear instability driving substantial deformation across a larger, if narrow, area and the proximity to several other voids within local void ensembles allowing direct coalescence..

We have shown here how the typical sequence of void nucleation, growth and coalescence can be imaged and interpreted at all relevant length scales above the individual site of damage nucleation that typically requires advanced imaging and local adjustment of imaging conditions, e.g. for electron channelling contrast imaging of the underlying martensite substructures and dislocations inside the ferrite [46, 47]. Together with the mechanistic understanding of damage initiation from the atomic scale of alloying and single phase or interface plasticity or decohesion, the quantitative as well as mechanistic understanding of damage evolution is directly relevant to formulating new models and informing existing simulation approaches with reliable data for calibration and validation of critical assumptions in the underlying damage models. Ultimately, it is the physical understanding across all length-scales that will allow us to formulate reliable damage models to guide microstructure and process design towards better material and component performance in terms of (light) weight, safety and cost or environmental impact. Having the right experimental methods at hand that allow the development and confident use of models and simulations is therefore an essential step in achieving this goal and connecting the scales of physical mechanisms and engineering design or industrial materials processing.

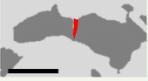

Figure 11: Schematic representation of the identified main causes for the three stages of damage evolution, both for the local evolution of a single void and for the global evolution of damage as an ensemble of voids. Panoramic high resolution micrographs and progress in automated single void damage analysis will improve our understanding of damage evolution and allow formulation and calibration of advanced models for damage simulation from the single phase to the sheet or component scale.

# Conclusions

We studied the influence of strain and triaxiality on the fundamental stages of deformation-induced void evolution (damage) by automated void detection and measurements on large-scale, high-resolution SEM images of differently notched tensile samples deformed to failure. While this approach globally confirmed the expected dependencies of void nucleation, growth and coalescence on strain and stress triaxiality, the multitude of observed and measured voids in our statistical approach highlighted the fact that an individual division of microstructural voids into clear, separate categories of the stages of damage evolution are not yet feasible. In detail, we conclude:

- The measured level of damage varies along with the critical amount of damage at the point of sample fracture: Higher values for triaxiality lead to a lower tolerance for damage at similar strains.
- While the evolution in terms of nucleation and initial growth of a single void is determined by the local stress state in the surrounding heterogeneous microstructure of DP800 steel, damage evolution across all voids converges against the expectations for the applied macroscopic stress state
- Globally, the applied plastic strain exerts a major influence on the overall number, and therefore the nucleation of new voids, while triaxiality correlates with the median void size and is then in turn related to the extent of void growth.
- Void coalescence in the large-scale images was detected predominantly along shear bands that introduce severe, local plasticity and allow voids to merge in areas of high void density, in particular martensite bands.
- In spite of the acquisition of high resolution data across a large area and use of recently developed first deep learning tools for automated mechanism analysis, a physical distinction of nucleation and growth remains elusive.
- No basis for the use of threshold void dimensions could be found to distinguish void nucleation and growth.

For a separation of these stages, a clear and physical definition of the onset of growth as well as a concurrent and automated analysis of mechanisms and geometry at all stages of void evolution will have to be achieved.

# Acknowledgments

The investigations are kindly supported by the German Research Foundation in context of the Collaborative Research Centre CRC/Transregio 188 "Damage-Controlled forming processes", projects B02 and B05, project number 278868966.